\documentclass[12pt,preprint]{aastex}
\usepackage{graphicx}
\usepackage{epsfig}
\usepackage{multirow}
\usepackage{footnote}

\shorttitle{Microwave Observations of Edge-on Disks}
\shortauthors{C. Melis et al.}

\begin{document}


\title{Microwave Observations of Edge-on Protoplanetary Disks: Program Overview and First Results}


\author{Carl Melis\altaffilmark{1,10}, G. Duch\^{e}ne\altaffilmark{2,3}, Laura Chomiuk\altaffilmark{4,5}, Patrick Palmer\altaffilmark{6}, M. D. Perrin\altaffilmark{7}, S. T. Maddison\altaffilmark{8}, F. M\'{e}nard\altaffilmark{3}, K. Stapelfeldt\altaffilmark{9}, C. Pinte\altaffilmark{3}, G. Duvert\altaffilmark{3}}
\email{cmelis@ucsd.edu}


\altaffiltext{1}{Center for Astrophysics and Space Sciences, University of California, San Diego, CA 92093-0424, USA}
\altaffiltext{2}{Astronomy Department, 601 Campbell Hall, University of California, Berkeley, CA 94720-3411, USA}
\altaffiltext{3}{UJF-Grenoble 1/ CNRS-INSU, Institut de Plan\'{e}tologie et d'Astrophysique de Grenoble (IPAG) UMR 5274, BP 53, 38041 Grenoble Cedex 9, France}
\altaffiltext{4}{Harvard-Smithsonian Center for Astrophysics, 60 Garden St., MS-66, Cambridge, MA 02138, USA}
\altaffiltext{5}{A  Jansky Fellow of the National Radio Astronomy Observatory}
\altaffiltext{6}{Department of Astronomy and Astrophysics, University of Chicago, 5640 S.\ Ellis Ave., Chicago, IL 60637, USA}
\altaffiltext{7}{Space Telescope Science Institute, 3700 San Martin Drive, Baltimore, MD 21231, USA}
\altaffiltext{8}{Centre for Astrophysics \& Supercomputing, Swinburne University, PO Box 218, Hawthorn, VIC 3122, Australia}
\altaffiltext{9}{Jet Propulsion Laboratory, California Institute of Technology, Mail Stop 183-900, 4800 Oak Grove Drive, Pasadena, CA 91109, USA}
\altaffiltext{10}{Joint NSF AAPF Fellow and CASS Postdoctoral Fellow}


\begin{abstract}
We are undertaking a multi-frequency Expanded Very Large Array (EVLA) survey
of edge-on protoplanetary disks
to probe the growth of solids in each disk, sedimentation of such material 
into the disk midplane, and the connection of these phenomena
to the planet formation process. 
The projection of edge-on disk systems along our line of sight enables a
study of the vertical stratification of large grains with fewer model dependencies
than would be required for disks that are more face-on.
Robust studies of the spatial distribution of grains up to $\approx$1\,cm in size
are possible with the wavelength range and sensitivity of the EVLA. 
In this contribution we
describe target selection and observational strategies. First results
concerning the Class~0 source IRAS\,04368+2557 (L1527\,IRS) 
are presented, including a study
of this source's 8.46\,GHz continuum variability over short and long time 
baselines and an indication that its protoplanetary disk 
may have a dearth of pebble-sized grains.
\end{abstract}





\section{Introduction}

Both major contending theories of planet formation, core accretion and gravitational
instability, require collection of solid material into a dense mid-plane layer within
circumstellar disks \citep[e.g.,][and references therein]{pollack96,boss97,schrapler04}. 
To understand this first stage in the generation of
planetary embryos we must determine the necessary conditions for
grain sedimentation and if the relevant physical processes 
act uniformly with particle size.
Numerical simulations incorporating gas drag and stellar gravity
predict that larger grains are expected to settle into the
disk mid-plane more efficiently than smaller grains \citep[e.g.,][]{barriere05,laibe08}.
However, these simulations
predict extremely short timescales for the
growth and migration of dust particles that are inconsistent with observations,
suggesting that some additional disk physics
needs to be included
(see \citealt{brauer07}).
Observational results, and modeling thereof 
\citep[e.g.,][]{duchene03,pinte07},
have begun to lay the foundation for grain sedimentation and its relation to grain
growth, providing evidence in support of larger grains being more concentrated towards
the disk mid-plane.
Although these works hint that grain growth, radial migration, 
and sedimentation are 
intimately connected, observations that resolve the spatial distribution of large
grains are necessary to complete this picture.

Study of the largest dust grains and their spatial distribution requires 
observations at long wavelengths (commensurate with grain size; e.g.,
\citealt{natta04,wilner05,rodmann06,lommen09}).
Unambiguous grain vertical distribution information can only come from 
protoplanetary disks which are
edge-on to our line of sight (Section \ref{secsamp}).
Edge-on protoplanetary disk systems imaged in scattered light typically
subtend $\sim$0.3-3$''$ in the vertical direction (see Table \ref{tabtargets} references). 
To identify vertical grain stratification,
thermal emission from disk atmosphere grains must be
separated from that of grains settled to the disk mid-plane with
high angular resolution observations.
Hence, to study the vertical distribution of large grains, one must map
edge-on disk systems with a long-baseline radio interferometer such as the
NRAO\footnote{The National Radio Astronomy Observatory is a facility of the National 
Science Foundation operated under cooperative agreement by Associated Universities, Inc.}
Expanded Very Large Array (EVLA; \citealt{perley11}).

In this contribution we present an EVLA survey of edge-on protoplanetary disk 
systems. Target selection and observations are discussed, and
first results on the source IRAS\,04368+2557 are reported.

\section{Sample Definition}
\label{secsamp}

Edge-on disks are selected because of their favorable geometry:
each beam of an edge-on disk map samples
a single disk altitude. From spectral index maps of edge-on disk systems
we will measure the maximum
grain size per synthesized beam
and hence the vertical distribution of grains
as a function of grain size. Some assumptions regarding disk axisymmetry may be
necessary to delineate optical depth effects from true grain size variation,
especially near the disk mid-plane
(see also discussion in Section \ref{secgg}).

We arbitrarily restrict our sample to disks with inclination angle 
$\gtrsim$75$^{\circ}$ (where 0$^{\circ}$ is face-on to our line of sight)
to limit confusion between radial and altitude flux variations.
All sources in our sample have 
been selected based on the existence of scattered light images that
allow a determination of the disk inclination to a few degrees. Ultimately, such
a sample will enable more powerful global panchromatic analyses
\citep[e.g.,][]{pinte08,duchene10}.

Disk systems observed in the first EVLA observing cycle are listed
in Table \ref{tabtargets}.


\section{Observational Strategy}

Preliminary observations aim to identify microwave-bright
disks for future mapping and
to characterize the overall degree of grain growth for each disk system based
on measurement of their long-wavelength opacity index $\beta$ (where
$\kappa_{\nu}$ $\propto$ $\nu$$^{\beta}$). The value of $\beta$ provides
information on the size of grains relative to the observing wavelength, 
where $\beta$$\lesssim$1
indicates grains comparable to or larger than the observing wavelength 
(e.g., \citealt{beckwith90,mannings94,rodmann06,lommen09,ricci10};
Section \ref{secgg} discusses how disk
optical depth affects the determination of $\beta$).
These goals require total power measurements (i.e., unresolved disk measurements)
which are best 
done in compact array configurations.

Disk flux measurements are made at 7 and 13\,mm.
Robust measurement of the opacity index $\beta$ requires
removing emission from processes other than
dust thermal emission.
Young stellar objects are known to emit in the microwave
due to free-free emission from ionized jets and disk-winds and
gyrosynchrotron emission from coronal processes (see e.g., \citealt{osten09}, and
references therein). Previous works have shown that free-free
and gyrosynchrotron emission (hereafter non-disk emission)
can contaminate the 7-13\,mm wavelength
region at a level that is anywhere
from 0-100\% of the detected flux \citep[e.g.,][]{natta04,rodmann06}.
It is assumed that measurements longward of 20\,mm are probing only
non-disk emission (this is not always the case; see \citealt{wilner05}).
We perform observations at both 35 and 60\,mm so that there are two
data points with which to determine the spectral slope of non-disk emission
components. 

Non-disk emission is known to be variable 
\citep[e.g.,][]{osten09}. Proper removal of
this variable emission requires observations of
the non-disk component that are as close in time as possible to observations
of the disk component. 
To obtain quasi-simultaneous observations across all bands,
we intertwined scans at 7, 13, 35, and 60\,mm within a continuous
observing block (which can span 3-5 hours; see Table \ref{tab04368}).
Although these disk and non-disk measurements are not perfectly simultaneous, 
the assumption is that averaging 35 and 60\,mm scans taken over the
entire observing block
will yield an accurate measurement of the non-disk emission component
during 7 and 13\,mm scans. We test this strategy in
Section \ref{secvar}.




\section{Case Study: IRAS\,04368+2557}

Of the 11 sources observed in the first EVLA cycle, 9 were detected
(Table \ref{tabtargets}). In this section observations of
IRAS\,04368+2557 (L1527\,IRS)
are presented as a case study of the methodologies outlined above. 

IRAS\,04368+2557 is an embedded class 0 object
in the L1527 dark cloud in Taurus \citep{white04}. Although optically
faint (e.g., \citealt{white04}), IRAS\,04368+2557 is bright in the
sub-millimeter \citep{chandler00,andrews05}, millimeter \citep{ohashi97,motte01},
and microwave \citep{rodriguez98,loinard02}. 
Millimeter and sub-millimeter results to date show that IRAS\,04368+2557
is composed of a substantial envelope that is infalling onto
the central source \citep{ohashi97,chandler00,motte01,andrews05}.
VLA imaging detects jet emission emanating from the central protostar
\citep{rodriguez98,reipurth04} and provides evidence of 
a $\sim$24\,AU binary companion \citep{loinard02}.
The disk surrounding the central source is probed most recently by
\citet{tobin10} who present 3.78\,$\mu$m imaging that resolves
the inner envelope and disk structure of 
IRAS\,04368+2557. The only unambiguous detection of disk thermal 
emission comes from \citet{loinard02} who resolve the disk at 7\,mm.
Our observations provide the first nearly simultaneous, multiband,
compact configuration radio frequency observations of IRAS\,04368+2557.

\subsection{Observations}

IRAS\,04368+2557 was observed with all 27 EVLA antennas. 
Some details of the observations are given in Table \ref{tab04368}. 
The WIDAR correlator was set up with two 128\,MHz sub-bands centered
on the frequencies listed in Table \ref{tab04368}. Each sub-band had 4 polarization
products (RR, LL, RL, LR) and sixty-four 2000\,kHz channels. Observations at
7 and 13\,mm were
performed in ``fast-switching'' mode; target source scans were interleaved
with frequent visits to a nearby calibration source
to freeze out rapid atmospheric phase
fluctuations. Cycle times of 2 and 4 minutes were used for 7 and 13\,mm, respectively.
The primary calibration source 3C286 was used to measure the 
complex bandpass and to set the absolute flux scale.

To probe the stability of the non-disk emission component over long and short time baselines
we retrieved VLA archival 35\,mm observations of IRAS\,04368+2557 (see Table 2).
To maintain homogeneity in this data set we only use observations performed 
with all VLA antennas. VLA observations were made in both circular polarizations
with an effective bandwidth of 92\,MHz centered at a frequency of 8.46\,GHz.
Data from AR0350 and AR0465 are presented in \citet{rodriguez98} and
\citet{reipurth04}, respectively. All other data are unpublished to the best
of our knowledge. While our  reductions agree well with those
of \citet{rodriguez98}, we find a significantly lower flux density for IRAS\,04368+2557
than do \citet{reipurth04}. Further investigation shows that our measurement
is consistent with what is displayed in Figure 2 of \citet{reipurth04}, but that
this and our measurement are both inconsistent with what is listed in their Table 2.

\begin{figure}
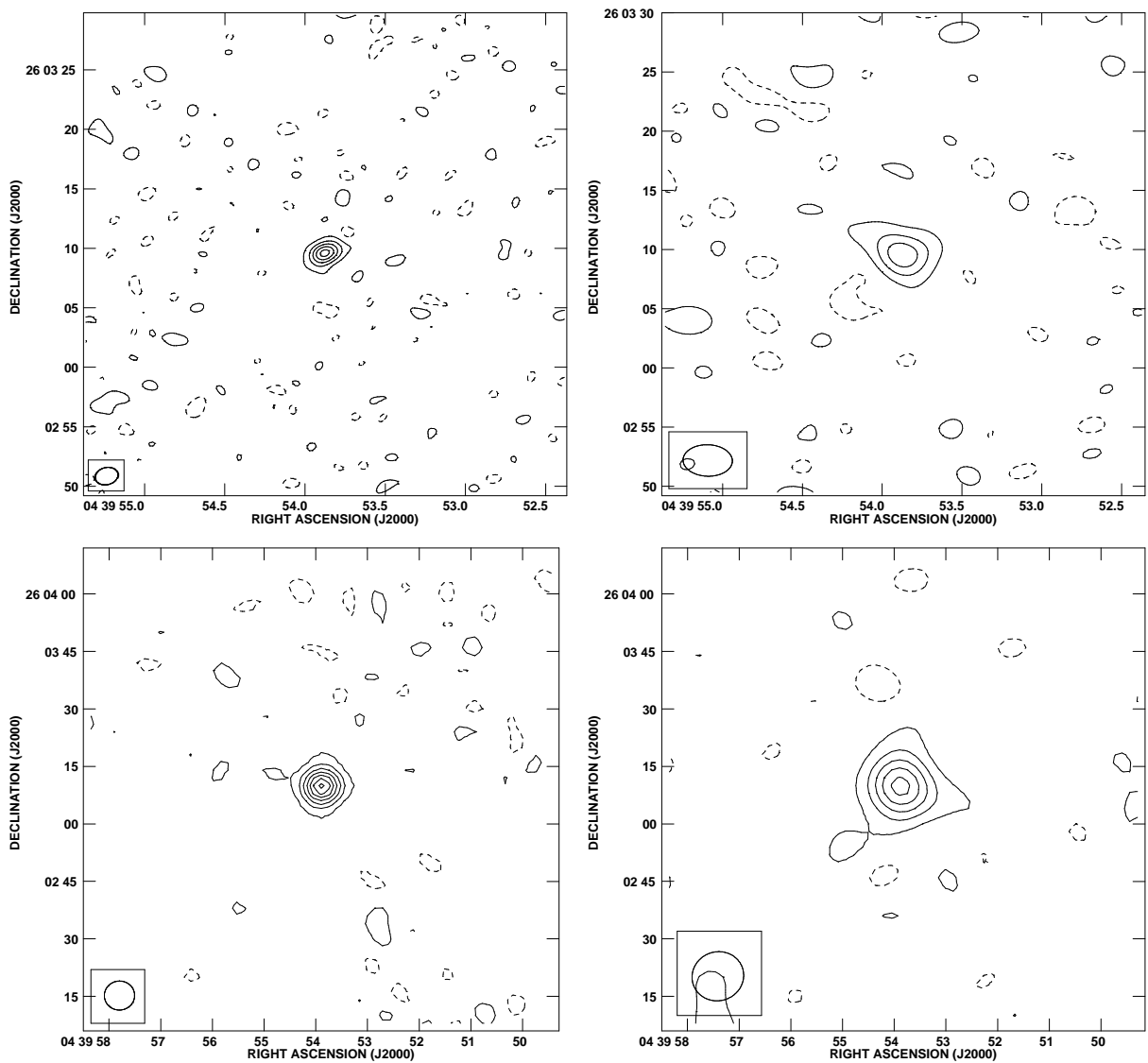

 \centering
 \begin{minipage}[t!]{80mm}
  \includegraphics[width=75mm,angle=-90]{figure1a.eps}
 \end{minipage}
 \begin{minipage}[t!]{80mm}
  \includegraphics[width=75mm,angle=-90]{figure1b.eps}
 \end{minipage}
 \\*
 \begin{minipage}[b!]{80mm}
  \includegraphics[width=75mm,angle=-90]{figure1c.eps}
 \end{minipage}
 \begin{minipage}[b!]{80mm}
  \includegraphics[width=75mm,angle=-90]{figure1d.eps}
 \end{minipage}
\caption{\label{figcntrs} EVLA images of IRAS\,04368+2557. The top row shows 7\,mm
               (left) and 13\,mm (right) maps while the bottom row shows 35\,mm (left) and
               60\,mm (right) maps. Half-power contours of the beam 
               are shown in the bottom left of each plot,
               beam sizes can be found in Table \ref{tab04368}. Contour levels start at $-$2 times
               the map rms noise level 
               (dashed contours) and increment by 4 times the map rms noise level (0.2 mJy, 
               0.1 mJy,
               30 $\mu$Jy, and 33 $\mu$Jy for 7, 13, 35, and 60\,mm respectively).}
\end{figure}

All data are reduced using the Astronomical Image Processing Software 
(AIPS; \citealt{greisen03}). VLA data are edited and calibrated following
standard VLA data reduction procedures. EVLA data are edited and calibrated in a similar
manner, except for the addition of first pass fringe fitting to set the interferometer
delays (performed on the primary calibrator, 3C286) and bandpass calibration. 
Standard high-frequency reduction techniques are employed for 7\,mm data.
IRAS\,04368+2557 is detected at all observed frequencies with $\gtrsim$10$\sigma$
significance. We assume absolute flux density scale systematic
uncertainties of 15\% and 5\% for 7 and 13\,mm, respectively; these are included
in the Table \ref{tab04368} uncertainties.
Absolute flux densities for 35 and 60\,mm  are assumed 
to be limited by the rms noise level in CLEANed maps.
Figure \ref{figcntrs} shows contour maps from the EVLA data; IRAS\,04368+2557 is
not resolved in any of the EVLA images.

\subsection{Centimetric Variability}
\label{secvar}

\begin{figure}
 \begin{minipage}[t!]{160mm}
  \includegraphics[width=160mm]{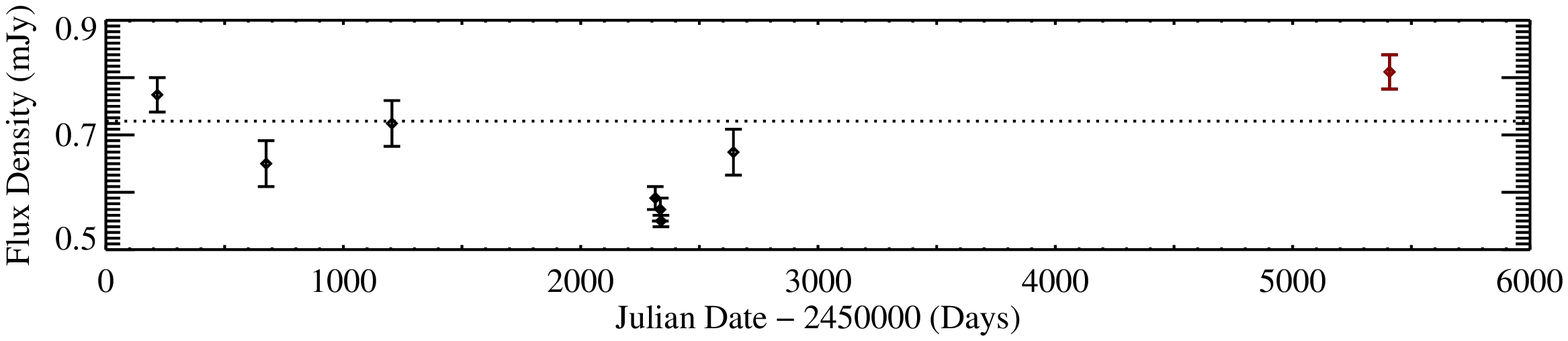}
 \end{minipage}
 \\*
 \begin{minipage}[t!]{80mm}
  \includegraphics[width=80mm]{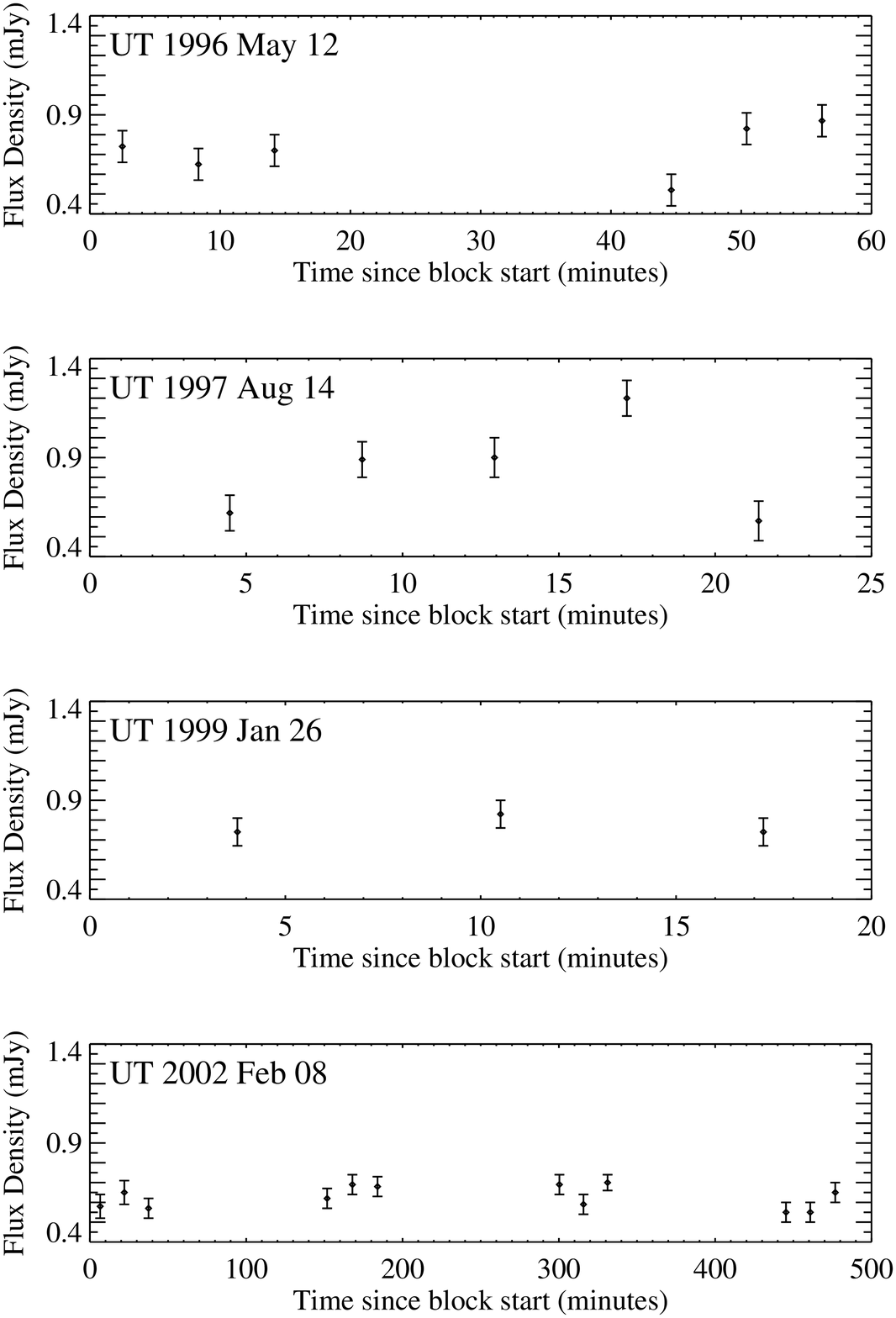}
 \end{minipage}
 \begin{minipage}[t!]{80mm}
  \includegraphics[width=80mm]{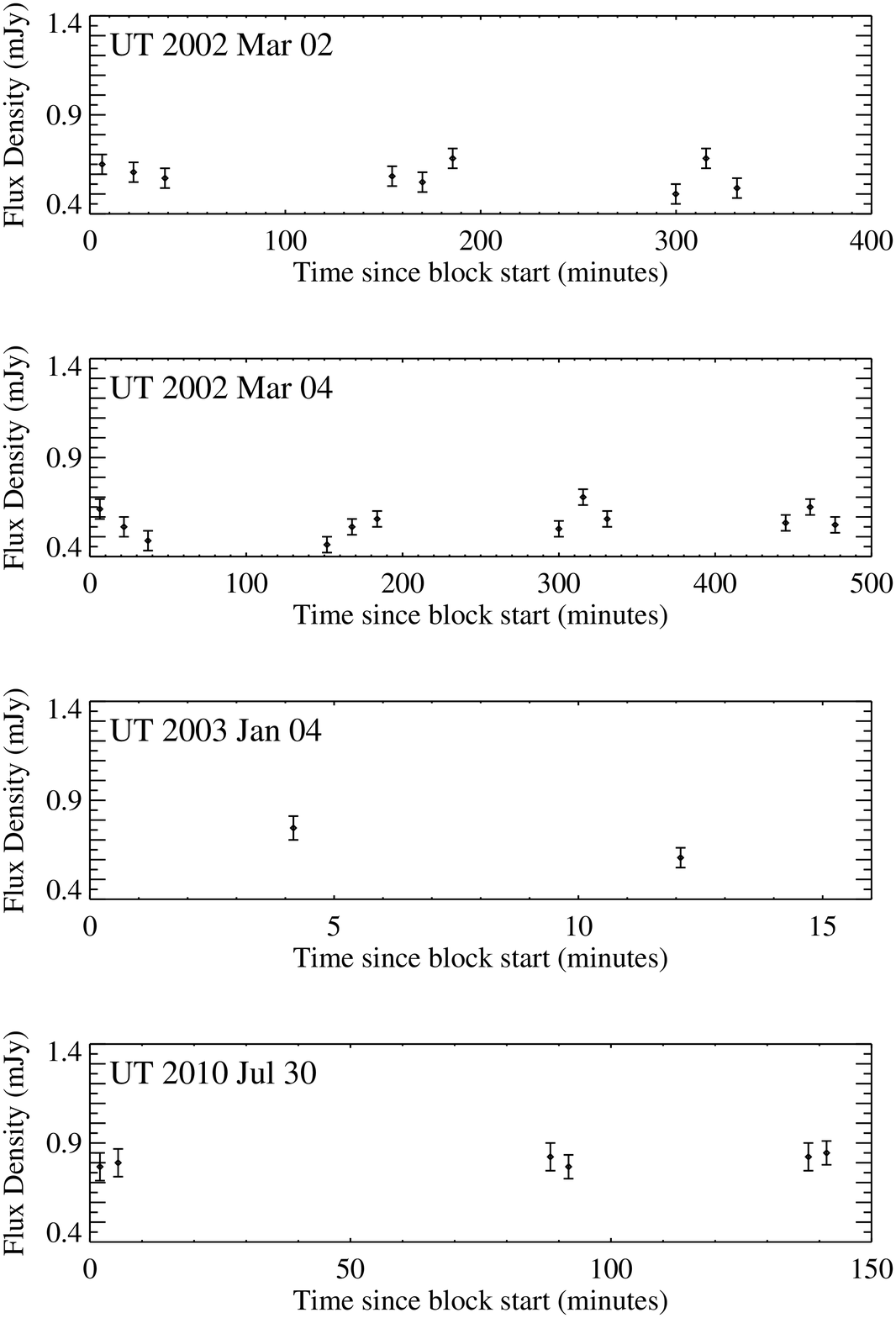}
 \end{minipage}
\caption{\label{figvar} X-band (8.46\,GHz; 35\,mm) measurements of IRAS\,04368+2557. The top
               panel presents inter-epoch measurements (one measurement for each date
               listed in Table \ref{tab04368}; EVLA data are in red). The dotted line is
               the mean flux level of all compact array measurements.
               The lower panels show intra-epoch measurements
               for the date listed in each plot. The abscissa shows time since the Table
               \ref{tab04368} UT start time of each epoch. 
               Note the mini-flare detected on UT 1997 Aug 14.}
\end{figure}

Figure \ref{figvar} presents 35\,mm measurements of IRAS\,04368+2557 from
each epoch listed in Table \ref{tab04368}. Inter-epoch measurements are made
by imaging all data obtained in the time interval listed for that 
epoch. Intra-epoch measurements are made
by imaging the smallest time interval that yields an $\approx$10$\sigma$
detection of IRAS\,04368+2557.

Except for the 2002 data, 
no significant ($>$3$\sigma$) inter-epoch variability is detected.
The 2002 observations were done with the most extended
array configuration and emission structures could have been resolved
out by the widely separated baselines resulting in lower than average
flux densities. Intra-epoch measurements reveal two transient events. 
The first is a mini-flare observed near the
end of the 26 minute long UT 1997 Aug 14 scan. 
The second is the radio jet detected by \citet{reipurth04} which
appears only near the end of the UT 2002 Feb 08 data set.
Out of eighteen $\approx$25 minute long 35\,mm 
IRAS\,04368+2557 sequences, one exhibits a flare event and one exhibits a
jet event. The occurrence rate of such events in 25 minute windows
is $\sim$6\% if the events are unique and $\sim$11\% 
if they are the same phenomenon.

Compared to previous studies of young stellar object centimetric variability
\citep[e.g.,][]{forbrich06,forbrich07,choi08,osten09},
the results presented herein appear to agree best with the \citet{osten09} 
study of short and long term variability in
six young stellar objects. At a wavelength of 60\,mm
they find that 4 out of 6 young stellar objects show short term 
variability and that 3 out of 6 objects show long term variability (the positive
variability detections are not necessarily from the same objects in each group).
Hence, their results suggest that young stellar objects
appear to be just as likely
to have short term variability as they are to exhibit long term variability.
Such conclusions seem to contradict the results of \citet{forbrich06,forbrich07} and
\citet{choi08}, where it is found that there are generally lower levels
of variability on shorter timescales. However, of the above
mentioned studies, only \citet{osten09} analyze
variability on intra-epoch (less than day) timescales. It
could be the case that even seemingly stable sources have sporadic variability
that can only be probed on the shortest timescales. If the event occurrence rate
derived above for IRAS\,04368+2557 is representative of stars in the same class,
then such sources might exhibit short-duration flares
in as many as one out of ten 25 minute observing windows.
However, since less than ten sources have been probed on 
intra-epoch timescales, it is likely premature to extend their results to other 
sources. Monitoring similar to that presented here and in \citet{osten09}
of new and previously studied young stellar objects
can further address variability timescales and strength in a more statistical sense.

For weak flare events like those detected here, averaging over
the duration of an observing block sufficiently suppresses
the flare effect. Stronger flares can likely be identified by analyzing
all data in a time-series fashion. No strong flares are present in the EVLA
data.

\subsection{Disk Emission Spectral Index}
\label{secgg}

\begin{figure}
 \includegraphics[width=160mm]{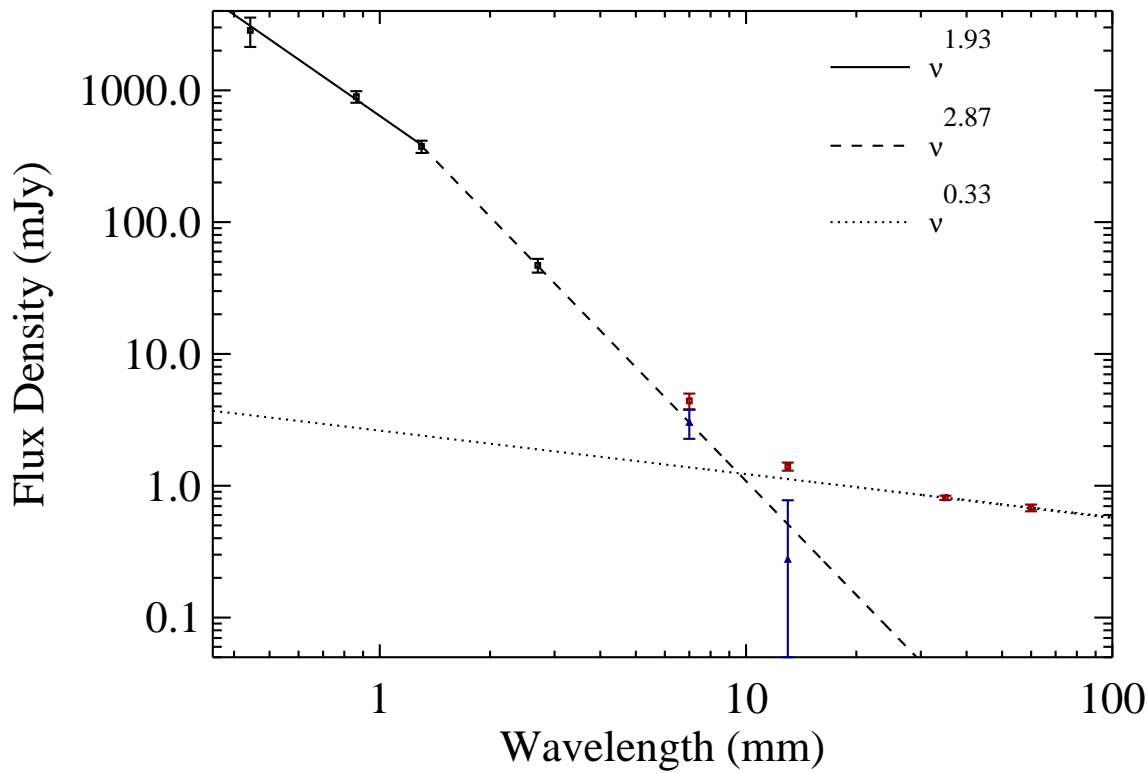}
\caption{\label{fig04368sed}
               IRAS\,04368+2557 long-wavelength spectral energy distribution.
               Uncertainties are 1$\sigma$.
               Red data points longward of 3\,mm are EVLA data.
               See Section \ref{secgg}
               and Table \ref{tab04368} for measurement beam sizes.
               The 450\,$\mu$m to 2.7\,mm measurement errors
               include 10-25\% absolute flux scale systematic uncertainties.
               The dotted line fits the
               35 and 60\,mm measurements which are assumed to be entirely due
               to non-disk emission (this fit does not include the 13\,mm data point). 
               Lower, blue triangle data points are corrected
               measurements (Section \ref{secgg});
               the 13\,mm uncertainty extends to 0. 
               The $\nu$$^{2.87}$ fit does not include
               the corrected 13\,mm measurement. See Section \ref{secgg} for a discussion
               of the two disk emission spectral indices used.}
\end{figure}

Figure \ref{fig04368sed} shows the EVLA flux measurements of 
IRAS\,04368+2557 with literature measurements at millimeter and sub-millimeter
wavelengths. 
The 450 and 850\,$\mu$m measurements \citep{andrews05} were made with 
beam sizes of 9 and 15$''$, respectively, while the
1.3\,mm measurement \citep{motte01} was made with a beam size
of 11$''$. The 2.7\,mm measurement \citep{ohashi97} was made with
a synthesized beam size of 6$''$ $\times$ 4.9$''$ (PA +163$^{\circ}$).

The non-disk emission component is fit with 
$\alpha$$_{\rm 35-60mm}$$=$$0.33\pm0.17$
(where $\alpha$ is the spectral index
F$_{\rm \nu}$ $\propto$ $\nu$$^{\alpha}$), 
consistent with non-disk indices assumed by \citet{rodmann06}.
An extrapolation of the non-disk emission fit is subtracted
from 7 and 13\,mm measurements;
uncertainties of each flux measurement and the spectral index are propagated
into the corrected flux uncertainties.
We calculate for IRAS\,04368+2557
$\alpha$$_{\rm 0.45-1mm}$$=$1.93$\pm$0.25
and $\alpha$$_{\rm 1-7mm}$$=$2.87$\pm$0.17
using the corrected 7\,mm flux density.
We use different spectral indices for wavelengths
$\lesssim$1\,mm and $\gtrsim$1\,mm since single power-law
fits to all sub-millimeter and millimeter data points are discrepant 
with the 450\,$\mu$m flux measurement at the $\approx$5$\sigma$ level.
The spectral index from 450\,$\mu$m to 1.3\,mm 
is consistent with Rayleigh-Jeans
emission suggesting that the disk is opaque at these wavelengths.
In the case of the 450\,$\mu$m to 2.7\,mm measurements,
there is likely contamination from extended envelope structure
detected around IRAS\,04368+2557
\citep[see below and][]{ohashi97,chandler00,motte01,andrews05}. 
If these measurements are contaminated by envelope 
emission, then the disk-only $\alpha$$_{\rm 1-7mm}$ is
smaller (flatter) than the value quoted above. It is noted that the EVLA
7\,mm flux measurement is roughly consistent with the total flux density estimated
for the resolved structures detected by \citet{loinard02}. This suggests that
the disk structure is fairly compact (relative to the envelope structure), 
in agreement with scattered light images
presented by \citet{tobin10}. IRAS\,04368+2557 is resolved at
wavelengths shorter than 7\,mm with beam sizes larger than the EVLA
7\,mm synthesized beam implying that those flux measurements
include envelope emission. Synthesis of the above
suggests that accurate measurement of
grain growth from spectral indices requires data sets having
beam sizes comparable to the disk angular size when contaminating
sources are present.

Extracting an opacity index $\beta$ from the above measured
$\alpha$$_{\rm 1-7mm}$ is complicated not only by envelope contamination
but also by edge-on disk geometry. When calculating $\beta$ one must
account for optically thick inner disk emission
sampled by the flux measurements. Should no optically thick
emission be present, then $\beta$ is simply 
$\alpha$$_{\rm 1-7mm}$$-$2. In the case that optically thick emission
is present in the measurements, $\beta$ takes the form
of $(1 + \Delta )$\,($\alpha_{\rm 1-7mm}$$-$2) where $\Delta$
is the ratio of optically thick to optically thin emission
\citep{beckwith90}. As discussed above, the disk orbiting
IRAS\,04368+2557 appears to be completely opaque out to wavelengths
as long as $\sim$1\,mm. Thus, significant optically thick emission may
be present even at wavelengths near 10\,mm. Estimates of
$\Delta$ are typically
made assuming some disk mass density profile and by measuring
the temperature profile from spectral indices and disk flux levels in
the optically thick wavelength regime \citep{beckwith90,rodmann06}. 
However, Equation 10 of \citet{beckwith90} shows that
mapping spectral indices and disk flux levels into
temperature profiles is confounded when 
disks are nearly edge-on. Modeling of edge-on disk
spectral energy distributions and resolved images can
recover their true disk temperature profiles.
Due to this additional modeling and envelope contamination uncertainties,
we leave derivation of the opacity index for the IRAS\,04368+2557 disk
to future works.

The 13\,mm measurement appears to fall short of the $\nu$$^{2.87}$ fit
after being corrected for non-disk emission.
This flux deficit suggests a break in the disk emission spectral index 
from 7 to 13\,mm, but is marginally significant.
If envelope contamination is present as discussed above,
then its removal would increase the significance of the deficient
corrected 13\,mm measurement.
Comparison to other protoplanetary disk systems shows that such a
break would be unusual (e.g., \citealt{wilner05,rodmann06,lommen09}) 
and suggestive of a cut-off near pebble-sizes in
the disk grain size distribution. If these characteristics
of IRAS\,04368+2557 are confirmed, then there may be a connection between
its potential $\sim$24\,AU binary companion (\citealt{loinard02}), 
its apparent 7\,mm outer disk truncation \citep{loinard02}, and
its possible lack of large disk grains.

\section{Conclusions}

We are carrying out an Expanded Very Large Array program to 
map the vertical distribution of large grains in edge-on
protoplanetary disks. The ultimate goal of this survey is to study
grain growth and sedimentation as the first stages of planet formation.
Our compact array observational strategy aims to provide
accurate disk thermal emission measurements in the microwave
by sampling non-disk emission in concert with disk measurements.
First EVLA results for IRAS\,04368+2557 show that the protoplanetary disk
around this source is likely optically thick out to millimeter wavelengths
and that it may have a dearth of ``pebble-sized'' grains.

\acknowledgments

We thank the anonymous referee for comments that helped improve this work.
C.M. acknowledges support from the National Science Foundation under
award No.\ AST-1003318.
M.D.P. was supported by NSF Postdoctoral Fellowship No.\ AST-0702933.
F.M., C.P., and G.D.\ acknowledge PNPS of CNRS/INSU, 
and ANR (contract ANR-07-BLAN-0221) of France for financial support.
C.P.\ acknowledges funding from the 
European Commission's 7$^\mathrm{th}$ Framework Program
(contract PIEF-GA-2008-220891) and from ANR under contract 
ANR-2010-JCJC-0504-01.



{\it Facilities:} \facility{EVLA ()}, \facility{VLA ()}

\clearpage





\clearpage

\clearpage

\clearpage

\begin{deluxetable}{lcccccccc}
\rotate
\tabletypesize{\normalsize}
\tablecolumns{9} 
\tablewidth{0pt}
\tablecaption{EVLA D-array Observed Targets \label{tabtargets}}
\tablehead{
 \colhead{Name} &
 \colhead{RA} &
 \colhead{DEC} &
 \colhead{Spectral} &
 \colhead{Disk} &
 \colhead{Inclination} & 
 \colhead{Disk Size\tablenotemark{a}} &
 \colhead{Ref} &
 \colhead{Detected?\tablenotemark{b}} \\
 \colhead{} &
 \multicolumn{2}{c}{(J2000 Phase Center)} &
 \colhead{Type} &
 \colhead{Class} &
 \colhead{($^{\circ}$)} &
 \colhead{($''$)} &
 \colhead{} &
 \colhead{}
}
\startdata
Haro 6$-$5B  & 04 22 01.0 & +26 57 35 & K5 & II & 75 & 4 & 1,2,3 & Y \\
IRAS\,04302+2247 & 04 33 16.5 & +22 53 20 & $-$ & I & 87 & 2 & 1,3,4 & Y \\
HV Tau C & 04 38 35.5 & +26 10 41 & M0 & II & 84 & 0.7 & 5,6,7 & Y \\
IRAS\,04368+2557 & 04 39 53.6 & +26 03 06 & $-$ & 0 & 85 & 1 & 1,8 & Y  \\
CB 26 & 04 59 50.7 & +52 04 44 & $-$ & I-II & 85 & 2.8 & 9 & Y \\
PDS 144 N\tablenotemark{c} & 15 49 15.4 & $-$26 00 52 & A2 & II & 83 & 0.8 & 10,11 & Y \\
LFAM1\tablenotemark{d} & 16 26 21.8 & $-$24 22 51 & $-$ & I & 85 & 1 & 12,13,14 & Y \\
Oph E MM3 & 16 27 05.9 & $-$24 37 08 & $-$ & II & 87 & 1 & 12,15 & Y \\
Flying Saucer & 16 28 13.2 & $-$24 31 39 & $-$ & II & 86 & 4.3 & 12,16 & N \\
Gomez's Hamburger & 18 09 13.4 & $-$32 10 50 & A0 & II & 84 & 12 & 17,18,19 & Y \\
HH 200 & 20 57 06.6 & +77 36 56 & $-$ & II & 87 & 1.5 & 20,21 & N  \\
\enddata
\tablerefs{(1) \citet{white04}, (2) \citet{krist98}, (3) \citet{stark06}, (4) \citet{wolf03}, (5) \citet{appenzeller05}, (6) \citet{stapelfeldt03}, (7) \citet{duchene10}, (8) \citet{tobin10}, (9) \citet{sauter09}, (10) \citet{vieira03}, (11) \citet{perrin06}, (12) \citet{vankempen09}, (13) \citet{bontemps01}, (14) \citet{duchene07}, (15) \citet{brandner00}, (16) \citet{grosso03}, (17) \citet{ruiz87}, (18) \citet{bujarrabal08}, (19) \citet{bujarrabal09}, (20) \citet{bally95}, (21) \citet{devine09}.}
\tablenotetext{a}{Disk angular diameter from optical scattered light or millimeter imaging. In its most extended array configuration the EVLA has synthesized beam sizes at 7 and 13\,mm of $\approx$0.043$''$ and 0.089$''$, respectively.}
\tablenotetext{b}{If source was detected in 7 and 13\,mm observations.}
\tablenotetext{c}{7\,mm observations with the VLA in the CnB array configuration; no 13\,mm observations were done.}
\tablenotetext{d}{Archival 7\,mm observations with the VLA, program AR0698; no 13 or 60\,mm observations were done.}
\end{deluxetable}

\clearpage

\begin{deluxetable}{lcccccccc}
\rotate
\tabletypesize{\normalsize}
\tablecolumns{9} 
\tablewidth{0pt}
\tablecaption{IRAS\,04368+2557 Observations \label{tab04368}}
\tablehead{
 \colhead{Project} &
 \colhead{Date} &
 \colhead{$\nu$} &
 \colhead{$\lambda$} &
 \colhead{Flux Density} &
 \colhead{Beam Size} &
 \colhead{Observation Interval} & 
 \colhead{TOS\tablenotemark{a}} &
 \colhead{Calibrator} \\
 \colhead{} &
 \colhead{(UT start)} &
 \colhead{(GHz)} &
 \colhead{(mm)} &
 \colhead{(mJy)} &
 \colhead{($''$)} &
 \colhead{(UT time)} &
 \colhead{(minutes)} &
 \colhead{}
}
\startdata
\multicolumn{9}{c}{EVLA Observations} \\
\tableline
AM1017 & 30 Jul 2010 & 43.3 & 7 & 4.4$\pm$0.6 & 1.96 $\times$ 1.44 (PA $-$76.1$^{\circ}$) & 11:56$-$13:32 & 22 & J0438+300 \\ 
AM1017 & 30 Jul 2010 & 22.5 &  13 & 1.4$\pm$0.1 & 4.16 $\times$ 2.66 (PA +88.3$^{\circ}$) & 11:41$-$13:16 & 26 & J0431+206 \\  
AM1017 & 30 Jul 2010 & 8.5 & 35 & 0.81$\pm$0.03 & 7.90 $\times$ 7.56 (PA $-$85.4$^{\circ}$) & 11:19$-$13:42 & 21 & J0431+206 \\
AM1017 & 30 Jul 2010 & 5.0 & 60 & 0.68$\pm$0.04 & 13.54 $\times$ 12.74 (PA $-$70.4$^{\circ}$) & 11:29$-$13:52 & 21 & J0431+206 \\
\tableline
\multicolumn{9}{c}{VLA Observations} \\
\tableline
AS0711 & 04 Jan 2003 & 8.5 & 35 & 0.67$\pm$0.04 & 2.45 $\times$ 2.18 (PA $-$51.3$^{\circ}$) & 06:48$-$07:04 & 16.17 & J0431+206 \\  
AR0465 & 04 Mar 2002 & 8.5 & 35 & 0.55$\pm$0.01 & 0.23 $\times$0.22 (PA +89.8$^{\circ}$)\tablenotemark{b} & 20:59$-$05:02 & 159.5 & J0403+260 \\  
AR0465 & 02 Mar 2002 & 8.5 & 35 & 0.57$\pm$0.02 & $-$ & 00:06$-$05:44 & 116.83 & J0403+260 \\
AR0465 & 08 Feb 2002 & 8.5 & 35 & 0.59$\pm$0.02 & $-$ & 23:03$-$07:06 & 156.5 & J0403+260 \\
AS0653 & 26 Jan 1999 & 8.5 & 35 & 0.72$\pm$0.04 & 2.72 $\times$ 2.44 (PA +70.7$^{\circ}$) & 06:29$-$06:50 & 20.5 & J0431+206 \\   
AE0112 & 14 Aug 1997 & 8.5 & 35 & 0.65$\pm$0.04 & 2.28 $\times$ 2.20 (PA +39.0$^{\circ}$) & 12:43$-$13:09 & 25.67 & J0431+206 \\ 
AR0350 & 12 May 1996 & 8.5 & 35 & 0.77$\pm$0.03 & 7.79 $\times$ 2.64 (PA $-$86.5$^{\circ}$) & 18:52$-$19:51 & 35.5 & J0403+260 \\ 
\enddata
\tablenotetext{a}{Time on source; aggregate of scan lengths only, does not account for slewing/settling of telescopes.}
\tablenotetext{b}{Synthesized beam size for all AR0465 data.}
\end{deluxetable}






\end{document}